\documentclass[prb,twocolumn,preprintnumbers,amsmath,amssymb]{revtex4}
\usepackage[dvips]{graphicx}
\usepackage{dcolumn}
\usepackage{bm}
\usepackage{times}
\usepackage{mathptmx}
\usepackage{color}
\usepackage{pifont}
\usepackage[latin1]{inputenc}
\usepackage{natbib}

\begin{document}

\title{Translation domains in multiferroics}

\author{D. Meier$^{\rm a}$, N. Leo$^{\rm a}$, T. Jungk$^{\rm b}$, E. Soergel$^{\rm b}$,
 P. Becker$^{\rm c}$, L. Bohat\'{y}$^{\rm c}$ and M. Fiebig$^{\rm a}$$^{\ast}$
\thanks{$^\ast$Corresponding author. Email: fiebig@hiskp.uni-bonn.de\vspace{6pt}}\\\vspace{6pt}
 $^{\rm a}${\em{HISKP, Universit\"at Bonn, 53115 Bonn, Germany}};
 $^{\rm b}${\em{PI, Universit\"at Bonn, 53115 Bonn, Germany}};
 $^{\rm c}${\em{Institut f\"ur Kristallographie, Universit\"at zu K\"oln, 50937 K\"oln, Germany}}  }

\begin{abstract}
Translation domains differing in the phase but not in the orientation of the corresponding order
parameter are resolved in two types of multiferroics. Hexagonal (h-) YMnO$_3$ is a
split-order-parameter multiferroic in which commensurate ferroelectric translation domains are
resolved by piezoresponse force microscopy whereas MnWO$_4$ is a joint-order-parameter
multiferroic in which incommensurate magnetic translation domains are observed by optical second
harmonic generation. The pronounced manifestation of the generally rather ``hidden'' translation
domains in these multiferroics and the associated drastic reduction of symmetry emphasize that the
presence of translation domains must not be neglected when discussing the physical properties and
functionalities of multiferroics.
\end{abstract}

\maketitle

\section{Introduction}

During the last decade an abundance of studies on systems with coexisting magnetic and electric
order, the so-called multiferroics, disclosed a variety of new and intriguing phenomena such as
``unconventional'' ferroelectricity or ``gigantic'' coupling effects between magnetic and electric
properties.\cite{Cheong07a,Tokura07a} The multifarious physics and the technological potential of
the multiferroics released a hunt for further members of this material family so that a steadily
increasing number of multiferroic compounds is now at our disposal. Irrespective of the diversity
of multiferroics and magnetoelectric coupling phenomena, they can be grouped into two fundamental
categories: In the \textit{split-order-parameter multiferroics} the electric and the magnetic
order emerge independently --- at separate transition temperatures and due to a different
microscopic origin. Hexagonal (h-) HoMnO$_3$ is a prominent example with $T_{\rm C}^{\rm
el}=875$~K and $T_{\rm N}^{\rm mag}=75$~K.\cite{Aken04a,Lottermoser04a} In the
\textit{joint-order-parameter multiferroics} the electric and the magnetic order have a common
microscopic origin. For instance, in Ni$_3$V$_2$O$_8$ the magnetic long-range order violates the
spatial inversion symmetry and induces a spontaneous polarization via the antisymmetric
Dzyaloshinskii-Moriya interaction.\cite{Lawes05a} Consequently, the symmetry-breaking magnetic and
electric order parameters arise at the same temperature ($T=6.3$~K).

An universal property of all multiferroics is a substantial reduction of the number of symmetry
elements by the multiple order which implies the formation of a large number of domains. In
particular, incommensurate systems have a lower symmetry compared to the non-ferroic
state because of the loss of the three-dimensional translation symmetry. With the unusual nature
of the magnetic and electric order in multiferroics and the additional degree of freedom arising
from their interaction, novel types of domains and domain-related effects are expected.
Unfortunately, we still have a very fragmentary understanding of the complex domain states in multiferroics --- the diversity of
multiferroic domains is largely ignored, although domains are the key to many macroscopic
properties of technological relevance such as coercive field, electrical resistivity, and
switching dynamics.

Investigations of domains in multiferroics so far are focussed on \textit{orientation domains},
i.e., domains differing in the orientation of magnetic and/or electric order
parameters.\cite{Fiebig02c,Zhao06a,Meier09b,Cabrera09a} In contrast, investigations of
\textit{translation domains}, i.e., domains differing in the phase but not the orientation of the
order parameter (so that they are also called anti-phase domains), are extremely rare because
their experimental observation is a very challenging task and their technological feasibility is
questionable.\cite{Meier09c} However, the substantial role the walls between translation
domains can play for magnetoelectric interactions was pointed out only
recently.\cite{Choi10a,Mostovoy10a}

In this paper we consider two complementary cases of translation domains in multiferroics that
demonstrate the extraordinary scope of manifestations: a split-order-parameter multiferroic with
commensurate order and a joint-order-parameter multiferroic with incommensurate order. Here,
h-YMnO$_3$ and MnWO$_4$ were chosen as model compounds because their magnetic and electric
properties are well understood. This is a good basis for studying the formation and properties of
their translation domains which are visualized using piezo-response force microscopy (PFM) and
optical second harmonic generation (SHG). We discuss the associated symmetry-breaking order
parameters and compare spatially resolved measurements of the domain topology in the two
compounds. Our analysis shows that they can be regarded as end cases in the classification of
translation domains in multiferroics. In h-YMnO$_3$ translation domains emerge due to
\textit{discrete} symmetry violations by a \textit{commensurate} order parameter strongly coupling
to the crystal lattice. In MnWO$_4$ translation domains emerge due to \textit{continuous} symmetry
violations by an \textit{incommensurate} order parameter. Here, the translation domains are
decoupled from the crystallographic structure.

\section{Order Parameters}

\subsection{Commensurate h-YMnO$_3$}

In h-YMnO$_3$ translation domains appear in conjunction with the ferroelectric order which emerges
in two steps:\cite{Lonkai04a,Fennie05a} A tripling of the paraelectric unit cell at 1350~K and
emergence of a spontaneous polarization $\pm P_z$ along the hexagonal axis at
1100~K.\cite{Nenert07a} The reduction of the space symmetry by the two transitions is $P6_3/mmc \;
\rightarrow \; P6_3cm$.

\begin{figure}
\centering
\includegraphics[clip,scale=0.6]{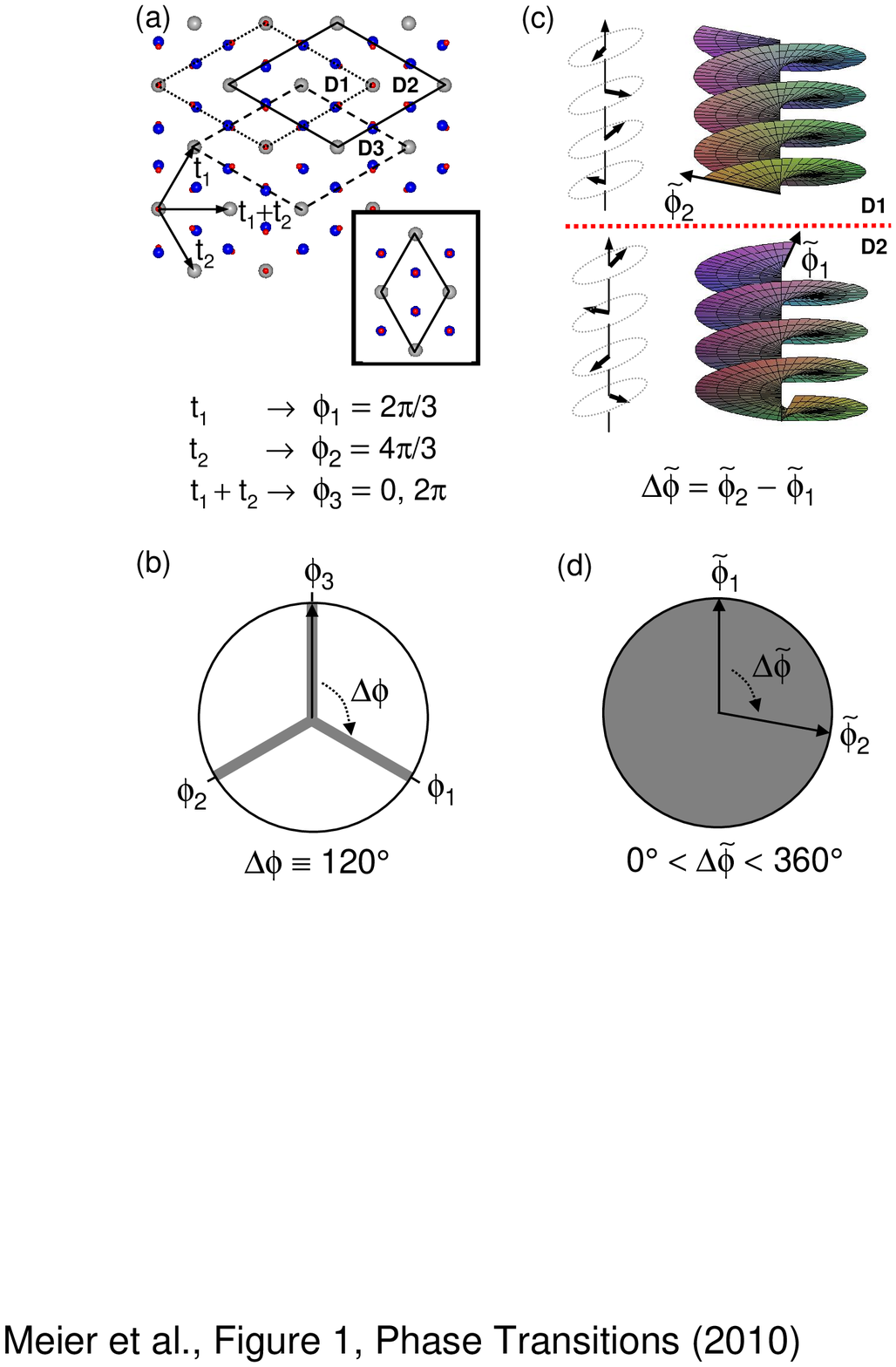}
\caption{\label{fig:Fig1} Origin of translation domains in h-YMnO$_3$ and MnWO$_4$. (a) Schematic
illustration of the three possible ferroelectric unit cells in h-YMnO$_3$ and the translations
$t_1$, $t_2$, and $t_1+t_2$ lost as symmetry operations with respect to the paraelectric unit cell
(inset). (b) Polar plot of the order-parameter space of the ferroelectric phase of h-YMnO$_3$. The
translations lost as symmetry operations have a discrete image in the order-parameter space so
that each translation domain can be identified by a specific phase $\phi_j$ ($j=1,2,3$). (c)
Schematic illustration of the elliptical spin spiral in the multiferroic phase of MnWO$_4$ (left)
and its dense continuation (right). (d) Polar plot of the order-parameter phase in the
multiferroic state of MnWO$_4$. Here, the translations lost as symmetry operations have a dense
image in the order-parameter space. Consequently, $\tilde\phi_j$ can have any value
$0^{\circ}<\tilde\phi_j<360^{\circ}$.}
\end{figure}
Because of the enlargement of the unit cell, which is sketched in Fig.~\ref{fig:Fig1}(a), the
translation symmetries $t_1$, $t_2$, and $t_3:=t_1+t_2$ of the high-temperature $P6_3/mmc$ phase
are violated. As a consequence, three different translation or trimerization domains denoted by
$\alpha$, $\beta$, and $\gamma$ arise,\cite{Choi10a} each being represented by one of the three
structurally nonequivalent unit cells depicted in Fig.~\ref{fig:Fig1}(a). The associated
two-dimensional order parameters are
\begin{equation} \label{comm}
\eta_j=\left(
    \begin {array} {c}
        \delta \mathrm{e}^{i\phi_j} \\
        \delta \mathrm{e}^{-i\phi_j} \\
    \end {array} \;
\right) \; (j=1,2,3)
\end{equation}
with $\mathbf k=(\frac{1}{3},\frac{1}{3},0)$ as commensurate wave vector describing the tripling.
The translation domains can be identified by their relative phase $\phi_j=\mathbf k\mathbf t_j$
with $\phi_1(\alpha)=\frac{2\pi}{3}$, $\phi_2(\beta)=\frac{4\pi}{3}$, and $\phi_3(\gamma)=0,
2\pi$. The values for $\phi_j$ are illustrated in the polar plot in Fig.~\ref{fig:Fig1}(b).
Their discreteness nicely reflects the commensurability of the translation domains in h-YMnO$_3$.
Apparently, all the translation domain walls in this compound correspond to a discontinuity in the
phase of the commensurate order parameter by $|\Delta\phi_{j\neq
j^{\prime}}|=|\phi_j-\phi_{j^{\prime}}|\equiv120^{\circ}$ with $j$ and $j^{\prime}$ denoting
different domains.

\subsection{Incommensurate MnWO$_4$}

As illustrated by Figs.~\ref{fig:Fig1}(c) and (d) the situation in MnWO$_4$ is quite different.
Here, translation domains evolve due to incommensurate magnetic long-range
order.\cite{Meier09a,Meier09c,Lautenschlaeger93a} First, a sinusoidal spin-density wave with the
propagation vector $\mathbf k=(-0.214,\frac{1}{2},0.457)$ emerges at 13.5~K. It is associated to a
single two-dimensional order parameter $\zeta_j$.\cite{Toledano10a} At 12.7~K the spin density
wave turns into an elliptical spin spiral with the same propagation vector as before. The transition
is driven by a second two-dimensional magnetic order parameter
\begin{equation} \label{inco}
\xi_j=\left(
    \begin {array} {c}
        \rho \mathrm{e}^{i\tilde{\phi}_j} \\
        \rho \mathrm{e}^{-i\tilde{\phi}_j} \\
    \end {array} \;
\right)  \; (j\in\mathbb{N}).
\end{equation}
Along with this, a magnetically induced polarization $P_y$ along the $y$ axis arises. Thus, the
system becomes multiferroic and orientation domains with $\pm P_y$ are formed. Note that it is the
\textit{coexistence} of the order parameters $\zeta_j$ and $\xi_j$, not the presence of $\xi_j$ by
itself, that breaks the inversion symmetry and induces a spontaneous polarization and, thus,
multiferroicity. In addition, $\zeta_j$ and $\xi_j$ give rise to translation domains that are
associated to the incommensurate spin arrangement in Figs.~\ref{fig:Fig1}(c).
Figure~\ref{fig:Fig1}(c) shows the elliptical spin spiral of the multiferroic phase in MnWO$_4$
and its dense continuation. The incommensurate magnetic structure described by $\mathbf k$ causes
a unit cell doubling along the $y$ axis and violates any translation symmetry in the $xz$
plane.\cite{Lautenschlaeger93a} In contrast to commensurate h-YMnO$_3$, the translation symmetries
violated by the incommensurate structure in MnWO$_4$ have a dense image in the order-parameter
space, see Figs.~\ref{fig:Fig1}(b) and \ref{fig:Fig1}(d). This gives rise to an infinite number
of translation domains enumerated by $j=1,2,3,\ldots$.\cite{Mettout10a} All translation domains
are energetically degenerated, because the free energy is invariant under the continuous rotation
of the phase mode (phason) associated with the incommensurate spin spiral. Note that the phase
between $\zeta_j$ and $\xi_j$ is unique so that it is sufficient to consider the order parameter
$\xi_j$ only for distinguishing and enumerating the translation domains.

In an intuitive interpretation one can say that the phase difference $\Delta\tilde{\phi}_{j\neq
j^{\prime}}=\tilde{\phi}_j-\tilde{\phi}_{j^{\prime}}$ appearing at a domain wall corresponds to a
discontinuity in the spatial precession of the incommensurate spin spiral. Any value
$\Delta\tilde{\phi}_{j\neq j^{\prime}}\in[0^{\circ},360^{\circ}]$ is possible due to the
incommensurate nature of the magnetic order. Note that the identification of a translation domain
is ambiguous. The phase $\tilde{\phi}$ changes continuously across a domain so that there is no
absolute criterion for labelling it. Only the \textit{phase difference} occurring at its boundary
defines a domain wall and, thus, distinguishes different domains.

In summary, two fundamentally different types of translation domains are present in h-YMnO$_3$ and
MnWO$_4$: In h-YMnO$_3$ a finite set of translation domains is generated by the commensurate order
parameter and a universal phase shift of $120^{\circ}$ distinguishes different domains. The
domains represent a pronounced structural discontinuity and are therefore expected to interact
strongly with the underlying crystal lattice and the ferroelectric displacement. In contrast,
MnWO$_4$ can develop an infinite number of translation domains caused by the continuous symmetry
breaking of the associated incommensurate magnetic order parameter so that the phase shift at the
domain wall can have any value $0^{\circ}<\Delta\tilde{\phi}<360^{\circ}$. Due to this continuous
nature only a weak coupling between the translation domains and the crystal lattice is expected.

\section{Domain Topology}

\subsection{Commensurate h-YMnO$_3$}

For investigating the translation domains in h-YMnO$_3$, flux-grown $z$-oriented platelets with a
lateral extension of a few millimeters and a thickness in the order of 100~$\mu$m were used. After
chemical-mechanical polishing with a silica slurry the distribution of domains was measured by
PFM.

\begin{figure}
\centering
\includegraphics{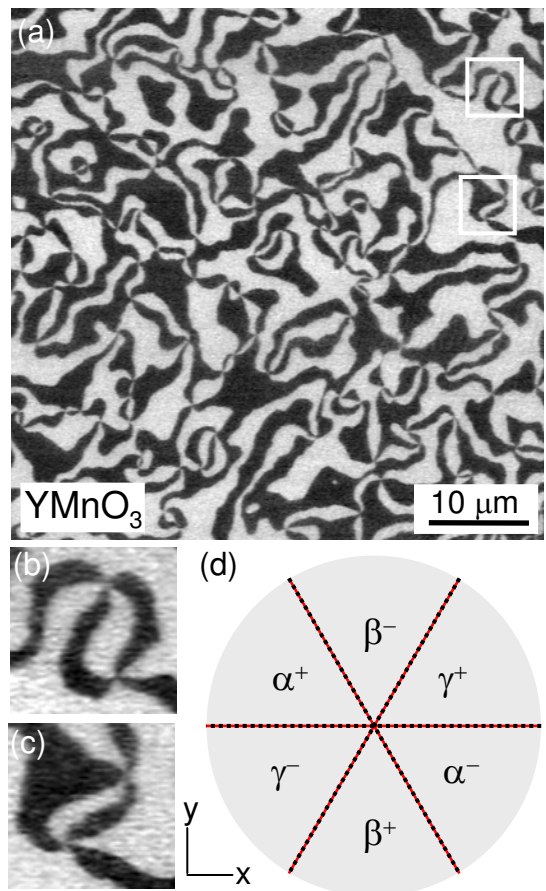}
\caption{\label{fig:Fig2}(a) PFM image of the $xy$ plane of an as-grown h-YMnO$_3$ crystal. Bright
and dark areas correspond to ferroelectric domains with $+P_z$ and $-P_z$, respectively. (b) and
(c) Enlargement of the domain intersections marked by boxes in panel (a). (d) Schematic of the
intersections. Note that only one type of domain wall is present because of the rigid coupling of
orientation and translation domains.}
\end{figure}

Figure~\ref{fig:Fig2}(a) shows the $xy$ plane of h-YMnO$_3$ under ambient conditions.
The PFM
image reveals domains of $\lesssim 1$~$\mu$m with two grey levels corresponding to ferroelectric
domains with the polarization pointing parallel $+P_z$ (bright) or antiparallel $-P_z$ (dark) to
the $z$ axis. The distribution of the domains is striking. All across the sample, a domain
structure with meeting points of six domains is obtained. No exceptions are detected --- whenever
the ferroelectric domains approach one another the characteristic kaleidoscopic pattern is formed.
The resulting topology is almost isotropic and can be observed in all three spatial
dimensions.\cite{Jungk10a}

The kaleidoscopic pattern with its meeting points of six domains can be understood on the basis of
Figs.~\ref{fig:Fig1}(a) and \ref{fig:Fig1}(b): In the ferroelectric phase each of the three
possible translation domains $\alpha$, $\beta$, and $\gamma$ can exhibit either a positive or a
negative spontaneous polarization, so that six different trimerization-polarization domains are
expected. As shown in Ref.~\onlinecite{Choi10a} meetings of different translation domains
\textit{require} a reversal of polarization in order to be stable. Therefore, when the three
possible translation domains meet in one point, this can only be solved by an arrangement of all
six translation-polarization domains. In Ref.~\onlinecite{Jungk10a} it was shown that the
arrangement has to be a sequence of the type shown in Fig.~\ref{fig:Fig2}(d) where the two order
parameters $\eta$ and $P_z$ of h-YMnO$_3$ change \textit{simultaneously} at each domain wall. For
instance, this allows an $\alpha^+, \beta^-, \gamma^+, \alpha^-, \beta^+, \gamma^-$ sequence of
domains around the meeting point, but it forbids an $\alpha^+, \alpha^-, \beta^+, \beta^-,
\gamma^+, \gamma^-$ arrangement. As explained in Ref.~\onlinecite{Jungk10a} the pairing of
orientation and translation domain walls is due to the electrostatic discontinuity occurring at
the translation domain walls. Hence, h-YMnO$_3$ is a remarkable example where translation domains,
in contrast to the common belief, manifest in a very pronounced way. In fact, the translation
domains \textit{control} the distribution of the orientation domains in h-YMnO$_3$ and with this
the distribution of the ferroelectric polarization. The electrostatic mechanism behind this
reflects the electrostatic, rather unusual nature of the ferroelectric order in h-YMnO$_3$ which,
in turn, is common to split-order multiferroics.

\subsection{Incommensurate MnWO$_4$}

We now turn to the complementary case, i.e., translation domains in an incommensurate
joint-order-parameter multiferroic, namely MnWO$_4$. The domain topology of MnWO$_4$ was
investigated by SHG since it is the only method allowing one to image antiferromagnetic
translation domains, based on its coupling to the phase of the magnetic order
parameter.\cite{Fiebig05a} SHG describes the induction of a dipole oscillation
$\mathbf{P}(2\omega)$ by an incident electromagnetic light wave given by $\mathbf{E}(\omega)$. It
is described by the equation $P_i(2\omega)=\epsilon_0\chi_{ijk}E_j(\omega)E_k(\omega)$ with $\hat{\chi}$
as nonlinear susceptibility of the material in which the optical frequency
doubling occurs. The set of components $\chi_{ijk}\neq 0$ reflects the symmetry and long-range
order of the material. They can couple to the orientation and phase of the order parameter and
thus be used to reveal the corresponding domain structure. In the following we will focus on the
discussion of the domain pattern observed in MnWO$_4$ whereas the discussion of the technical
details of the SHG in MnWO$_4$ are found elsewhere.\cite{Meier09a,Meier09b,Meier09c,Meier10a}

\begin{figure}
\centering
\includegraphics[width=0.45\textwidth]{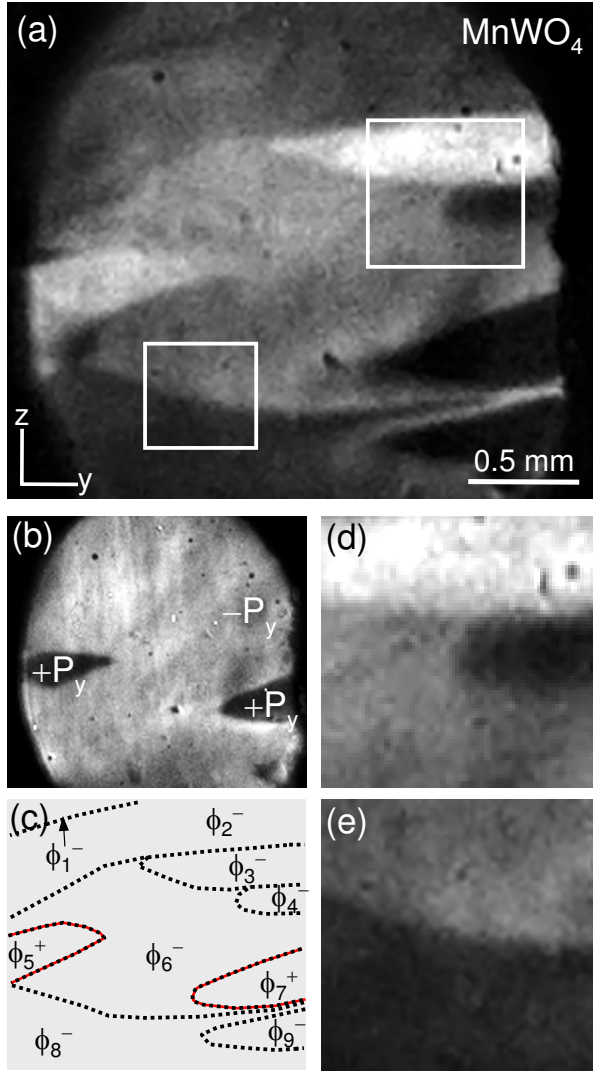}
\caption{\label{fig:Fig3}(a) Coexistence of orientation and translation domains in the $yz$ plane of
zero-field-cooled MnWO$_4$. The SHG image was gained in the multiferroic phase at 8~K and shows a
mixture of translation-polarization domain walls and translation-only domain walls. (b)
Distribution of the ferroelectric orientation domains. (c) Schematic of the image in (a). Red
solid lines denote the distribution of orientation domains of opposite polarization ($+P_y$ or
$-P_y$). Black dotted lines separate different translation domains labelled by $\phi_j$,
$(j=1-9)$. (d) and (e) Magnified meeting points of domains.}
\end{figure}

For the SHG measurements melt-grown MnWO$_4$ samples were prepared in the same way as the
h-YMnO$_3$ samples and measured in a transmission geometry.\cite{Becker07,Fiebig05a} The samples
had a lateral extension of $2.6\times3.2$~mm and a thickness of 830~$\mu$m. Figure~\ref{fig:Fig3}
displays a SHG image of the $yz$ plane exposed in the multiferroic phase at 8~K.
The image reveals
differently shaded regions separated by abrupt changes of the SHG intensity. These regions
correspond to different multiferroic domains. The different brightness is caused by the dependence
of the amplitude and phase of the SHG contribution on the relative orientation and phase of the
order parameters $\zeta$ and $\xi$.\cite{Fiebig05a} Thus, Fig.~\ref{fig:Fig3}(a) displays both
orientation \textit{and} translation domains. For disentangling them, the wavelength and
polarization of the incident light in Fig.~\ref{fig:Fig3}(b) were chosen such that only the
ferroelectric orientation domains distinguished by their spontaneous polarization $\pm P_y$ were
probed. This reveals two degrees of brightness corresponding to domains with $+P_y$ and $-P_y$. By
comparing Figs.~\ref{fig:Fig3}(a) and (b) two different types of domain walls in MnWO$_4$ can be
distinguished: (i) polarization-translation walls, and (ii) translation-only walls.

(i) At the domain walls marked by the red dotted lines in Fig.~\ref{fig:Fig3}(c) the orientation
of $P_y$ is reversed \textit{and} a phase discontinuity of $\tilde\phi$ in the complex order
parameter of Equation~\ref{inco} occurs. The latter can be concluded because the change of
brightness occurring upon crossing the polarization wall is not unique so that aside from the
spontaneous polarization another aspect of the order must have changed --- here the phase
$\tilde\phi$ is the only option.

(ii) Figure~\ref{fig:Fig3}(a) reveals additional domain structures within the polarization domain
denoted by $-P_y$. Here, the abrupt changes in the SHG yield correspond to translation-only walls
at which only the phase $\tilde\phi$ of $\xi$ changes. The domains are distinguishable by SHG
because of the aforementioned correspondence between the phase shifts of the SHG wave and the
order parameter $\xi$.\cite{Meier09c,Meier10a} As sketched in Fig.~\ref{fig:Fig3}(c) more than
six translation domains are resolved in Fig.~\ref{fig:Fig3}(a). This observation is consistent
with the effective continuous symmetry breaking by the associated incommensurate order parameter
giving rise to an unlimited number of translation domains.

In order to point out peculiar features of the topology of the translation domains
Figs.~\ref{fig:Fig3}(d) and (e) show magnifications of regions revealing, respectively, a
meeting point of three translation domains and a line separating two translation domains. The two
SHG images indicate that the topology of the incommensurate translation domains is not subject to
such rigid constraints as in the case of the commensurate split-order-parameter multiferroic
h-YMnO$_3$. In the joint-order-parameter multiferroic MnWO$_4$ the number of translation domains
that can meet in one point is not limited from the point of view of symmetry. The evolution of an
arbitrary number of translation domains is a specific property of incommensurate systems, clearly
separating them from commensurate matter, where a well-defined number of translation domains
emerges in the ordered state. In addition, translation domains in incommensurate systems expand
the established concept of translation domains by involving translations \textit{exceeding} the
expansion of the unit cell. Finally, as pointed out earlier,\cite{Mettout10a} incommensurabilities
do not only increase the number of allowed domain states --- they can also have a strong impact on
the thermodynamic features related to the incommensurate phase transition.

\section{Conclusion}

A violation of the three-dimensional translation symmetry by magnetic, electrical, or structural
degrees of freedom gives rise to different types of translation in multiferroics. In contrast to
orientation domains with a unique orientation of the corresponding order parameter the translation
domains were largely ignored in literature so far because of their ``hidden'' nature. Here we
demonstrated two end cases on the range of translation domains: commensurate ferroelectric
translation domains in the split-order-parameter multiferroic h-YMnO$_3$ and incommensurate
magnetic translation domains in the joint-order-parameter multiferroic MnWO$_4$. The type of
translation domain and the type of multiferroicity in these two compounds are intricately related.
However, two basic scenarios become obvious: Translation symmetries can either be broken by
\textit{commensurate} or \textit{incommensurate} long-range order with fundamental consequences
regarding the number of possible domain states and the formation of domain walls.

Our results show that in spite of their generally subtle nature, translation domains can be
associated to pronounced structural discontinuities. Consequently, the formation of translation
domains must not be neglected when discussing the physical properties of multiferroics or other
materials allowing the formation of translation domains. Moreover, the low symmetry at the site of
the domain walls can be the origin of new correlation effects. Here, the manipulation of
translation domains may even provide a handle for controlling the functionality of materials.

\textbf{Acknowledgments:} This work was supported by the DFG through the SFB 608 and the Deutsche Telekom AG. The authors
thank Pi\`erre Tol\'edano for many fruitful discussions.

\bibliographystyle{unsrt}

\begin{thebibliography}{21}

\bibitem{Cheong07a}
S.-W.~Cheong and M.~Mostovoy, {\em Multiferroics: a magnetic twist for ferroelectricity}, Nature Mat. 6 (2007), pp. 13 -- 20.

\bibitem{Tokura07a}
Y.~Tokura, {\em Multiferroics -- toward strong coupling between magnetization and polarization in a solid}, J. Magn. Magn. Mater. 310 (2007), pp. 1145 -- 1150.

\bibitem{Aken04a}
B.B.~van Aken, T.T.M.~Palstra, A.~Filippetti, and N.A.~Spaldin, {\em The origin of ferroelectricity in magnetoelectric YMnO$_3$}, Nature Mat. 3 (2004), pp. 164 -- 170.

\bibitem{Lottermoser04a}
Th.~Lottermoser, Th.~Lonkai, U.~Amann, D.~Hohlwein, J.~Ihringer, and M.~Fiebig, {\em Magnetic phase control by an electric field}, Nature 430 (2004), pp. 541 -- 544.

\bibitem{Lawes05a}
G.~Lawes et al., {\em Magnetically driven ferroelectric order in {Ni$_3$V$_2$O$_8$}}, Phys. Rev. Lett. 95 (2005), 087205.

\bibitem{Fiebig02c}
M.~Fiebig, Th.~Lottermoser, D.~Fr\"ohlich, A.V.~Goltsev, and R.V.~Pisarev, {\em Observation of coupled magnetic and electric domains}, Nature 419 (2002), pp. 818 -- 820.

\bibitem{Zhao06a}
T.~Zhao et al., {\em Electrical control of antiferromagnetic domains in multiferroic BiFeO$_3$ films at room temperature}, Nature Mat. 5 (2006), pp. 823 -- 829.

\bibitem{Meier09b}
D.~Meier, N.~Leo, M.~Maringer, Th.~Lottermoser, P.~Becker, L.~Bohat\'y, and M.~Fiebig, {\em Topology and manipulation of multiferroic domains in {MnWO$_4$}}, Phys. Rev. B 80 (2009), 224420.

\bibitem{Cabrera09a}
I.~Cabrera et al., {\em Coupled Magnetic and Ferroelectric Domains in Multiferroic Ni$_3$V$_2$O$_8$}, Phys. Rev. Lett. 103 (2009), 087201.

\bibitem{Meier09c}
D.~Meier, N.~Leo, Th.~Lottermoser, P.~Becker, L.~Bohat\'y, and M.~Fiebig, {\em Imaging of hybrid-multiferroic and translation domains in a spin-spiral ferroelectric}, Mater. Res. Soc. Symp. Proc. 1199 (2010), 1199-F02-07.

\bibitem{Choi10a}
T.~Choi, Y.~Horibe, H.T.~Yi, Y.J.~Choi, W.~Wu, and S.-W.~Cheong, {\em Insulating interlocked ferroelectric and structural antiphase domain walls in multiferroic YMnO$_3$}, Nature Mat. 9 (2010), pp. 253 -- 257.

\bibitem{Mostovoy10a}
M.~Mostovoy, {\em A whirlwind of opportunitues}, Nature Mat. 9 (2010), pp. 188 -- 190.

\bibitem{Lonkai04a}
Th.~Lonkai, D.G.~Tomuta, U.~Amann, J.~Ihringer, R.W.A.~Hendrikx, D.M.~T\"obbens, and J.A.~Mydosh, {\em Development of the high-temperature phase of hexagonal manganites}, Phys. Rev. B 69 (2004), 134108.

\bibitem{Fennie05a}
C.J.~Fennie and K.M.~Rabe, {\em Ferroelectric transition in YMnO$_3$ from first principles}, Phys. Rev. B 72 (2005), 100103(R).

\bibitem{Nenert07a}
G.~N\'enert, M.~Pollet, S.~Marinel, G.R.~Blake, A.~Meetsma, and T.T.M.~Palstra, {\em Experimental evidence for an intermediate phase in the multiferroic YMnO$_3$}, J. Phys.: Condens. Matter 19 (2007), 466212.

\bibitem{Meier09a}
D.~Meier, M.~Maringer, Th.~Lottermoser, P.~Becker, L.~Bohat\'y, and M.~Fiebig, {\em Observation and coupling of domains in a spin-spiral multiferroic}, Phys. Rev. Lett. 102 (2009), 107202.

\bibitem{Lautenschlaeger93a}
G.~Lautenschl\"ager, H.~Weitzel, T.~Vogt, R.~Hock, A.~B\"ohm, M.~Bonnet, and H.~Fuess, {\em Magnetic phase transitions of {MnWO$_4$} studied by the use of neutron diffraction}, Phys. Rev. B 48 (1993), pp. 6087 -- 6098.

\bibitem{Toledano10a}
P.~Tol\'edano, B.~Mettout, W.~Schranz, and G.~Krexner, {\em Directional magnetoelectric effects in MnWO$_4$: magnetic sources of the electric polarization}, J. Phys.: Condens. Matter 22 (2010), 065901.

\bibitem{Mettout10a}
B.~Mettout, P. Tol\'edano, and M. Fiebig, {\em Symmetry replication and toroidic effects in the multiferroic pyroxene NaFeSi$_2$O$_6$}, Phys. Rev. B 81 (2010), 214417.

\bibitem{Jungk10a}
T.~Jungk, \'A.~Hoffmann, M.~Fiebig, and E.~Soergel, {\em Electrostatic topology of ferroelectric domains in YMnO$_3$}, Appl. Phys. Lett. 97 (2010), 012904.

\bibitem{Fiebig05a}
M.~Fiebig, V.V. Pavlov, and R.V. Pisarev, {\em Second-harmonic generation as a tool for studying electronic and magnetic strucutres of crystals}, J. Opt. Soc. Am. B 22 (2005), pp. 96 -- 118.

\bibitem{Meier10a}
D.~Meier, N.~Leo, G.~Yuan, Th.~Lottermoser, P.~Becker, L.~Bohat\'y, and M.~Fiebig, {\em Second harmonic generation on incommensurate structures: The case of MnWO$_4$}, preprint: arXiv:1008.2864 (2010).

\bibitem{Becker07}
P.~Becker, L.~Bohatý, H.~J. Eichler, H.~Rhee, and A.~A. Kaminskii, {\em High-gain raman induced multiple stokes and anti-stokes generation in monoclinic multiferroic MnWO$_4$ single crystals}, Laser Phys. Lett. 4 (2007), pp. 884 -- 889.


\end{thebibliography}

\end{document}